# Millimeter-scale layered MoSe$_2$ grown on sapphire and evidence for negative magnetoresistance

M.T. Dau,[1,2,3] C. Vergnaud,[1,2,3] A. Marty,[1,2,3] F. Rortais,[1,2,3] C. Beigné,[1,2,3] H. Boukari,[1,4] E. Bellet-Amalric,[1,5] V. Guigoz,[1,6] O. Renault,[1,6] C. Alvarez,[1,7] H. Okuno,[1,7] P. Pochet,[1,7] and M. Jamet[1,2,3]

[1] Université Grenoble Alpes, F-38000 Grenoble, France

[2] CEA, INAC-SPINTEC, F-38000 Grenoble, France

[3] CNRS, INAC-SPINTEC, F-38000 Grenoble, France

[4] CNRS, Institut NEEL, F-38000 Grenoble, France

[5] CEA, INAC-PHELIQS, F-38000 Grenoble, France

[6] CEA, LETI, Minatec Campus, F-38054 Grenoble, France

[7] CEA, INAC-MEM, F-38000 Grenoble, France

**Abstract**

Molecular beam epitaxy technique has been used to deposit a single layer and a bilayer of MoSe$_2$ on sapphire. Extensive characterizations including in-situ and ex-situ measurements show that the layered MoSe$_2$ grows in a scalable manner on the substrate and reveals characteristics of a stoichiometric 2H-phase. The layered MoSe$_2$ exhibits polycrystalline features with domains separated by defects and boundaries. Temperature and magnetic field dependent resistivity measurements unveil a carrier hopping character described within two-dimensional variable range hopping mechanism. Moreover, a negative magnetoresistance was observed, stressing a fascinating feature of the charge transport under the application of a magnetic field in the layered MoSe$_2$ system. This negative magnetoresistance observed at millimeter-scale is similar to that observed recently at room temperature inWS2 flakes at a micrometer scale [Zhang et al., Appl. Phys. Lett. 108, 153114 (2016)]. This scalability highlights the fact that the underlying physical mechanism is intrinsic to these two-dimensional materials and occurs at very short scale.

**I. INTRODUCTION**

Layer-structured transition metal dichalcogenides (TMDs) have drawn much attention recently since they are being considered as a new class of semiconducting two-dimensional (2D) materials with thickness-tunable band-gap modulation [1–3]. In the single-layer configuration, the TMDs offer a unique platform to explore not only the carrier transport in an ultrathin channel but also the control of 2D excitonic systems and the spin-valley physics [4–6]. These may be achieved owing to their electronic structure and spatial-inversion symmetry that give rise to a strong spin-obit coupling, i.e. a large spin-splitting of the valence band, and to a versatile control of the spin and valley properties of the carriers [7,8].

Chemical vapor deposition has been employed for fabrication of TMDs layers. This technique enables to a large coverage of species on substrates [9–11]. However, the TMDs layers obtained by this method show an inhomogeneity and discontinuous nature due to the absence of precise thickness control. Molecular beam epitaxy (MBE) is of particular interest because one can achieve high purity of the layers, precise control of the thickness, flexible choice of metals and scalability [12,13]. Since a single layer of TMDs is separated from each other by an out-of-plane van der Waals gap, interactions between TMDs layers and a chemically passivated substrate are expected to be of van der Waals type. In this regard, kinetic energies of deposited atoms are in agreement with the dynamism of van der Waals assembly by employing MBE technique.

Here, we report the achievement of MBE $MoSe_2$ layers on sapphire-c(0001) combined with electrical demonstration. $MoSe_2$, which belongs to TMDs family, exhibits abundant functionalities related to its atomically thin crystal and electronic structure, leading to a wide range of applications in electronics, optoelectronics, valley-spintronics and photocatalysis [1,14,15]. Very recently, $MoSe_2$ gained more attention and has been claimed as a potential candidate for room temperature exciton-polaritonic devices [16]. In this letter, we focus on growth and characterizations of single layer $MoSe_2$ and on electrical properties of bilayer grown at 550 °C. In-plane X-ray diffraction (XRD) with respect to the azimuths of the substrate shows that the layer does not have any preferential in-plane orientation. Polycrystalline character of the layer results from disoriented domains those size is estimated about 8.0 nm. A clear spatial separation of 6.5 Å between the layer and the substrate, which corresponds to a van der Waals gap, was evidenced by cross-sectional scanning transition electron microscopy (STEM). The layer was found to be stoichiometric based on X-ray photoemission spectroscopy (XPS) and its homogeneity was demonstrated by using Raman spectroscopy. The electrical measurements as a function of temperature and magnetic fields were performed on the bilayer $MoSe_2$ with the contacts separated by milimeter-scaled distance. The carrier transport exhibits a clear temperature dependence which is well described by the two-dimensional Mott's variable range hopping (VRH) model. Interestingly, a decrease in resistance, i.e. negative magnetoresistance (MR), was found up to room temperature when an external magnetic field was applied. With the transport study in $WS_2$ flakes [17], the observation of negative MR in the MBE $MoSe_2$ clearly points to intrinsic feature of transport properties of 2D TMDs which is extended down to very small scale.

## II. RESULTS AND DISCUSSION

Reflection high-energy electron diffraction (RHEED) patterns and intensity profiles of the post-annealed monolayer are shown in Fig. 1(a). The RHEED diagrams are streaky and remain the same when rotating the samples around the growth axis. That indicates a good crystalline quality of the surface and a presence of multi-domains having no preferential in-plane orientation. Full width at half maximum (FWHM) of $MoSe_2$ $(0\bar{1})$ streak normalized over streak-to-streak distance is plotted and displayed in Fig. 1(b). It can be seen that the layer grown at 550 °C has a smallest FWHM, indicating a largest coherent size of domains in this layer compared to those of other layers. The structural properties of $MoSe_2$ were investigated by in-plane XRD (Fig. 1(c)). The reflections of the $MoSe_2$ monolayer grown at 550 °C are identified as (100), (110) and (200), and are consistent with a 2H-$MoSe_2$ phase [18]. The extracted in-plane lattice parameter is 3.28 ± 0.01 Å, which is identical to the bulk value. This is a strong indication of a van der Waals epitaxy which permits a structural adaptation despite a large misfit (31.3%) between the substrate and the layer. The presence of both (100) and (110) in-plane reflections of the $MoSe_2$ along both the (h00) and (hh0) directions of the substrate is in agreement with the polycrystalline nature of the layer as observed in RHEED patterns. From the broadening of the $MoSe_2$ peaks taken in Fig. 1(c) and using the Scherrer equation, one can estimate the size of the monodomains to about 8 nm.

Such a small domain size implies the presence of many structural defects, boundaries in a continuous layer.

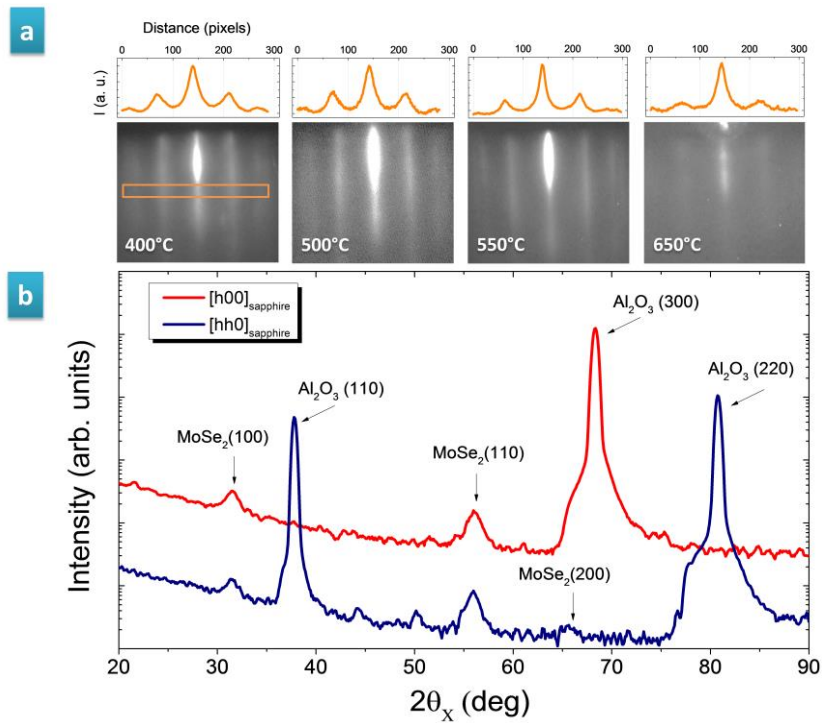

FIG. 1. (a) RHEED patterns of post-annealed samples grown at temperatures from 400 °C to 650 °C. The corresponding intensity profiles are shown on the top panel of the patterns. (b) FWHM of the (0$\bar{1}$) streaks taken from (a) normalized by the distance between (0$\bar{1}$) and (00) streaks as a function of growth temperature. (c) In-plane XRD of the monolayer MoSe$_2$ grown at 550 °C along two principal reflections (h00) and (hh0) of the substrate.

Figure 2 displays the cross-sectional STEM micrograph of the MoSe$_2$/sapphire interface. The bright line, which is parallel to the surface of the substrate, corresponds to the monolayer of MoSe$_2$. It can be seen that the interface between the layer and the sapphire is abrupt with a separation of about 6.5 Å wide as for the interlayer van der Waals gap in MoSe$_2$. This underlines the expected weak chemical coupling between the MoSe$_2$ and the substrate, resulting in a quasi-free-standing monolayer of MoSe$_2$ on sapphire. The STEM image also reveals a disrupted region (dotted arrow in Fig. 2(b)) where defects or domain edges might be located and they supposedly result from domain merging, Mo depletion or Se vacancies. Finally, one clearly distinguishes a regular bright point network at the topmost layer of the substrate. This seems to be of different chemical nature from the substrate due to different contrasts. Interfacial interactions involving oxygen of the first layer of the substrate and deposited adatoms could be a possible reason for the formation of this buffer layer.

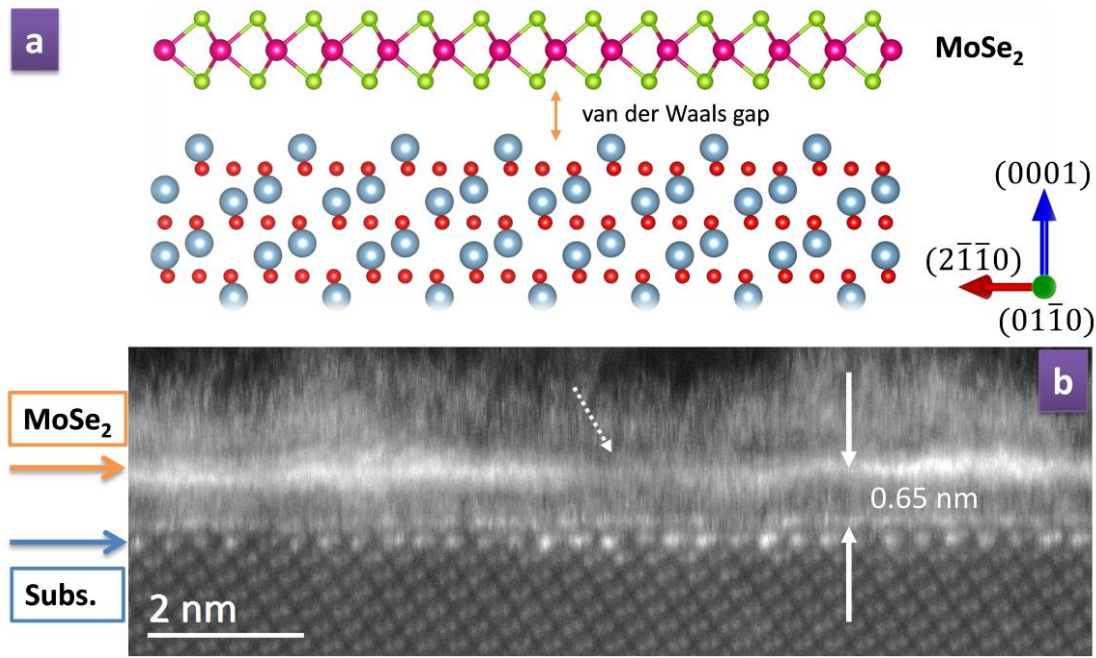

FIG. 2. (a) Ball-and-stick model represents the interface between the single layer and the substrate. (b) Cross-sectional STEM image of the MoSe$_2$/sapphire interface. The white arrows indicate the van der Waals gap and the dotted arrow indicates a disruption of the single layer.

High-resolution Mo3d and Se3d XPS spectra are shown in Fig. 3(a) and 3(b). The Mo3d peak fitting highlights a main doublet (component I) with the 3d$_{5/2}$ peak located at 229.2 ± 0.1 eV binding energy, and is assigned to stoichiometric MoSe$_2$: indeed, the energy is in agreement with the value of 229.3 eV reported for bulk MoSe$_2$ [19]. The energy separation of 174.4 eV for the main Se3d peak is also consistent with stoichiometric MoSe$_2$ systems [19,20]. In the Mo3d spectrum, a second, much weaker doublet, denoted as component II, is shifted by 1 eV to higher binding energies and is most likely due Mo sub-oxides [21]. Oxidized Se species are also found with the weak Se3d component II. Besides the Se3s line, the Mo3d spectrum fitting also evidences a third doublet measured at 233.7 eV (3d$_{5/2}$ binding energy) and assigned to MoO$_3$ species. We attribute these states to the interfacial layer as observed from STEM. We determined the stoichiometry of the MoSe$_2$ layer by considering the area of I and II Mo3d components, the area of the Se3d peak and the suitable relative sensitivity factors: by this we infer an overall composition of the layer of MoSe$_{1.97}$, which confirms the excellent quality of the sample, close to that of bulk MoSe$_2$.

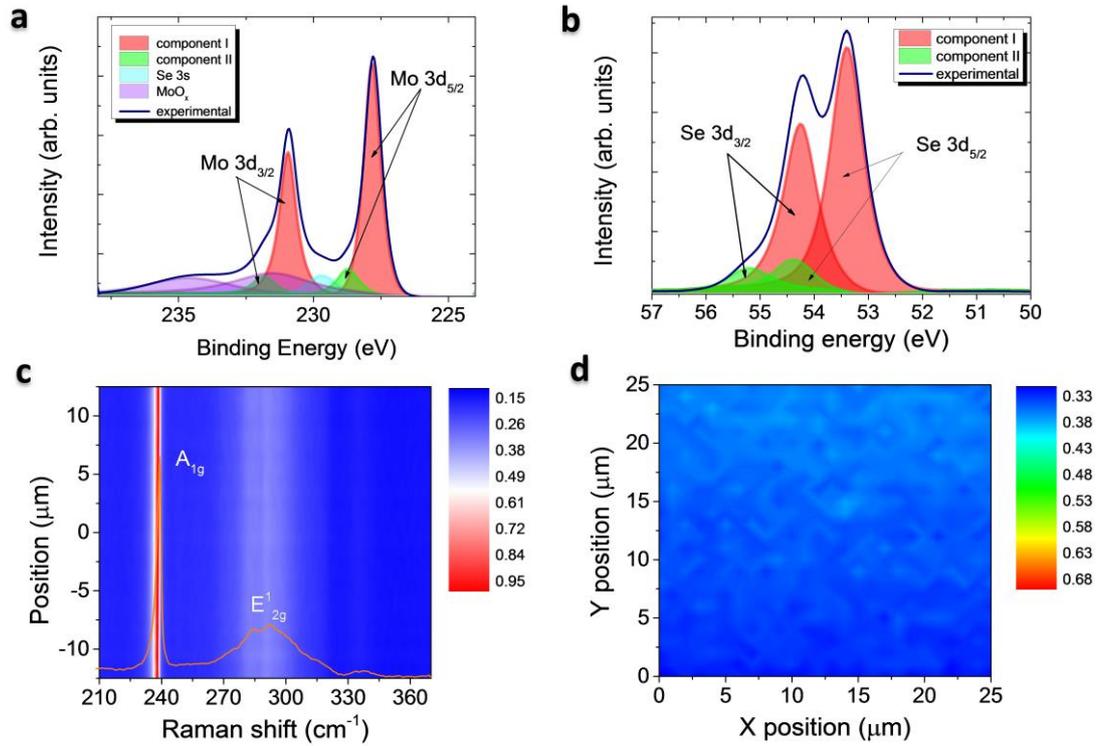

FIG. 3. (a) and (b) are respectively, Mo3d and Se3d XPS core-level spectra. (c) Position-dependent Raman shift in the 210-360 cm$^{-1}$ region. The intensity was normalized by A$_g^1$ peak intensity. (d) Intensity map of the E$_{2g}^1$/A$_g^1$ ratio with a standard deviation $\approx$ 0.012.

The Raman results are presented in Fig. 3(c) and 3(d). In Fig. 3(c), the main contributions are assigned to the first-order Raman peaks A$_g^1$ and E$_{2g}^1$ , which correspond respectively to the out-of-plane and in-plane vibration modes of 2H- MoSe$_2$. The Raman shifts are 241.1 cm$^{-1}$ for A$_g^1$ and 289.7 cm$^{-1}$ for E$_{2g}^1$. These values are in good agreement to the ones reported for bulk and exfoliated MoSe$_2$ [22]. The E$_{2g}^1$ peak is significantly broadened. This might be related to the structural defects in the layer including Se vacancies and domains [23] or to multi-peaks arising from the activation of the second-order Raman processes [24]. A much weaker peak is also detected around 338 cm$^{-1}$ and attributed to multi-phonon scattering [24]. It is known that the breathing mode B$_{2g}^1$ can be either Raman or an infrared active due to the spatial symmetry of the TMDs [25,26]. This mode is Raman active for few monolayers and the corresponding peak appears at around 352 cm$^{-1}$, whereas it becomes Raman inactive for one monolayer. For our monolayer MoSe$_2$, the absence of B1 2g mode in the spectrum confirms that the layer consists of one-layer sheet within the spatial resolution of the laser spot. Additionally, the two characteristic peaks $A_g^1$ and $E_{2g}^1$ of MoSe$_2$, are homogenously observed and their position does not shift wherever the acquisition takes place. The intensity ratio of these two peaks, which is used as an indicator of a homogeneous thickness, is also mapped over a size of 25x25 µm$^2$ and displayed in Fig. 3(d). The mapping shows an homogeneity of the layer and thoroughly corroborates the macroscopic observations. It is worth noting that the measurements were carried out at different positions of the sample and showed similar results.

The structural and spectroscopic properties of one monolayer MoSe$_2$ on sapphire were found to be comparable to that of bulk. The relevant attributes of the layer, that are polycrystalline feature with disoriented domains and nanoscopic disruption, were evidenced by analyzing XRD and STEM. Owing to the disruption, the current injection could not be achieved in the monolayer. To study the transport with a macroscopic contacts length, we grew a bilayer MoSe$_2$ to make sure continuity in the layer. The carrier transport was investigated with four points geometry illustrated in Fig. 4(b). The cross-sectional STEM image (Fig. 4(a)) shows that the interface between Mo/Pd and MoSe$_2$ is continuous and smooth. Fig. 4(c) displays a plot of temperature-dependent resistivity and the corresponding conductivity. An insulating characteristic of the layer with exponential dependence of resistivity was found for the studied range of temperature. As discussed previously for the single layer MoSe$_2$, the bilayer is also polycrystalline (isotropic RHEED patterns), hence, it is expected to be composed of disoriented domains with high density of vacancies and boundary defects. The structural defects could give rise to localized states that are in-gap electronic states and located near the Fermi level as revealed by density functional theory calculations and scanning tunneling spectroscopy in layered-TMDs systems [9,27]. In such a strongly disordered layer, VRH is expected to be dominant among transport mechanisms [28]. The logarithm of conductivity is also plotted (Fig. 4(c)) and found to be linearly dependent on $T^{-1/3}$, which is indeed in good agreement with the VRH transport theory [28]. Note that the fit was made by assuming that the MoSe$_2$ bilayer is of two-dimensional nature. The conductivity can therefore be expressed as: $\sigma \sim exp[-(T_0/T^{1/3})]$ where $T_0$ is the Mott characteristic temperature. The value of $T_0$ obtained by the fit is 1.1x10$^6$ K. This high value, which is comparable to the one recently reported in a thick layer of MoSe$_2$ [29], implies a strong localization of carriers mediated by the localized states. Furthermore, the Mott characteristic temperature in 2D system can be written: $T_0 = 13.8/(N_\mu k_B \xi^2)$ where $N_\mu$ is density of states near the Fermi level, $k_B$ is Boltzmann constant and $\xi$ is localization length [30]. By assuming temperature-independent $N_\mu = g_v m^*/(\pi \hbar^2)$ = 4.77 x $10^{14}$ eV$^{-1}$ cm$^{-2}$ with in-plane effective mass $m^*$=0. 57$m_0$ and valley degeneracy $g_v$ = 2 for MoSe$_2$ [31], $\xi$ can be written: $\xi = [13.8/(N_\mu k_B T_0)]^{1/2}$ = 1.75 Å . The value $\xi$ is in the same range order of lattice spacing and much smaller than domain size (8 nm), confirming a strong localization of states and the dominance of vacancy defects-driven transport.

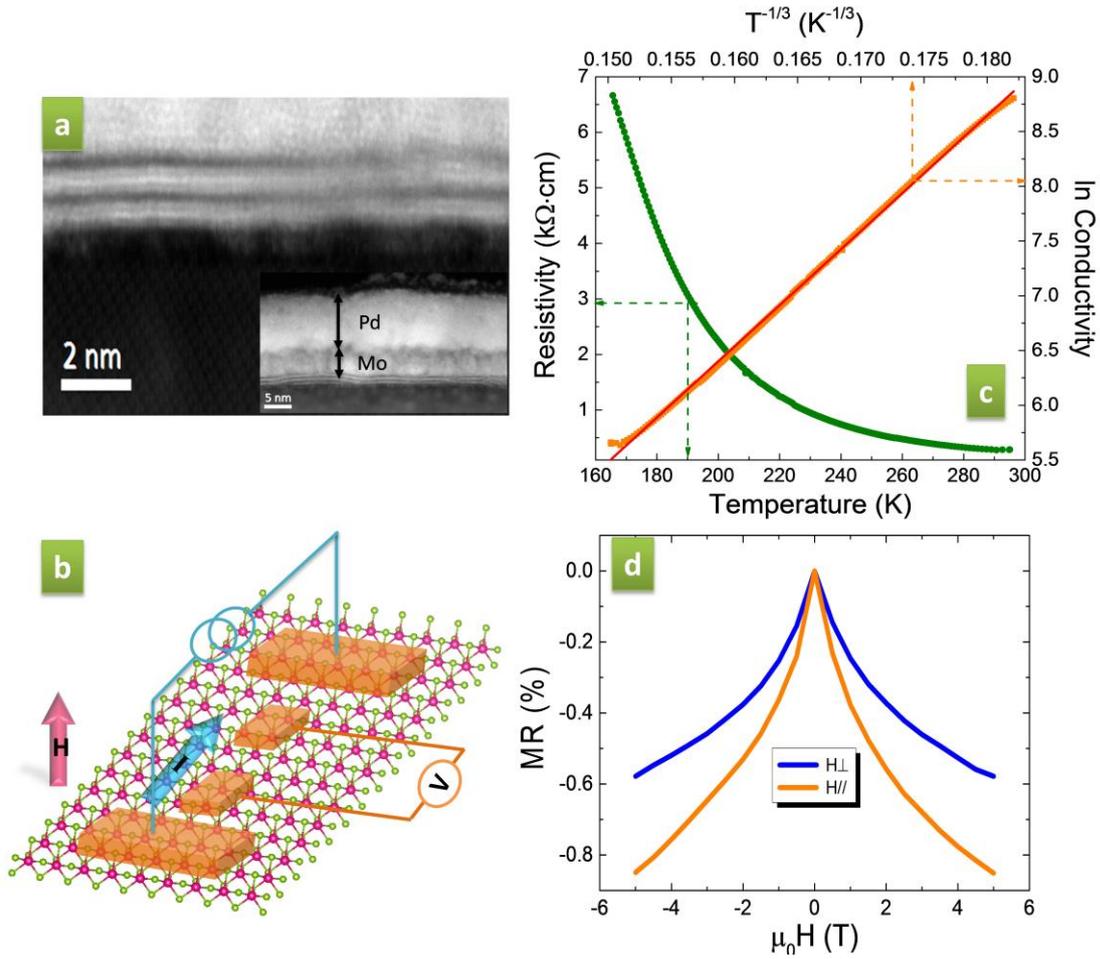

FIG. 4. (a) Cross-sectional STEM images of the bilayer MoSe$_2$/sapphire interface and the entire stack Mo/Pd/MoSe$_2$/sapphire (the inset). (b) Sketch of the measurement set-up with I the electrical current, V the bias voltage and H the applied magnetic field. (c) Carrier conductivity against temperature and the logarithm of the conductivity against T$^{-1/3}$ fitted with the 2D-VRH model (solid red line). (d) Magnetic field dependence of the magnetoresistance measured at 250 K with an external magnetic field applied parallel (H$_{//}$) and perpendicular (H$_\perp$) to the sample surface.

The magnetoresistance, *MR(%)=[R(T)-R(0)]/R(0)*, was found to be negative over the whole range of magnetic fields. The representing data obtained at 250 K are shown in Fig. 4(d). One can see that the MR is negative throughout the magnetic field range and does not show any indication of saturation at $\mu_0 H$ = 5 T. We note that the negative MR was also observed while the magnetic field was applied parallel to the sample surface, indicating that the negative variation of the resistance is related as well to the anisotropic orbital effect. Earlier reports on the magneto-transport properties of highly defected graphene [32, 33] and of exfoliated multilayers TMDs $WS_2$ [17] pointed out the negative MR in layered system. The finding of negative MR in bilayer $MoSe_2$ clearly stresses the pivotal implication of defects in the magnetic field dependent carrier transport. In our bilayer, the negative MR, which has never been observed so far for MBE or CVD TMDs, likely results from the hopping character of the VRH transport at high temperatures where the quantum corrections may not set in due to the predominance of inelastic scattering events. In a system exhibiting a low charge carrier density and a high degree of disorder, the hopping rate between localized states writes: $P \sim exp\left(\frac{2R}{\xi} - \frac{1}{\pi R^2 N_\mu k_B T}\right)$, where *R* is the hopping length [34]. The optimal hopping length corresponding to a maximum hopping rate is deduced: $R_0 = \left(\frac{\xi}{\pi N_\mu k_B T}\right)^{1/3}$. The maximum hopping rate *P* is therefore proportional to localization length $\xi$. Consequently, the decrease of resistance seems to result from a change of $\xi$ due to magnetic field. Negative MR in VRH regime was theoretically studied and found linear or quadratic in magnetic field on the basis of the hopping lengths $\xi$ and *$R_0$* [35]. The linear negative MR of bilayer $MoSe_2$ in low magnetic fields can be simply described by these models where the hopping is associated with interference of various paths. In the moderate-field region, MR is linearly dependent on magnetic field and is given: $MR \sim R_0^{3/2} \xi^{1/2} B = f(\xi) T^{-1/2} B$. The slope values obtained with the MR curves in the range of 0.05-0.75 T at different temperatures were obtained and then plotted versus temperature (see Supplementary Material). With the data in hand, the linear fit of ln(*slope-MR*) *vs* ln(*T*) yields -0.46 ± 0.12 which is somewhat different from -0.5. That is probably due to a significant dependence of the localization length on the magnetic field as we aforementioned. It is noted that a possible contribution on negative MR stems from polaron-type interactions of local magnetic moments coupled with localized carriers. Spin polarization occurred in metal orbitals due to a large density of Se vacancies would enable long-range coupling of magnetic polarons with an application of magnetic fields [36]. This could increase the carrier hopping rate, thus giving rise to the decrease in resistance of the layer. For the studied range of temperatures, this kind of interaction is probably weak in MoSe$_2$ layer due to thermal fluctuations and phonon scattering. Finally, the origin of the negative MR related to an enhancement of mean free path for carriers scattering at boundaries can be ruled out in our layer for the following reason. The domain size of MoSe$_2$ layer is estimated about 8 nm which is in the length range where boundaries might scatter the charge carriers over a long distance as a result of the enhanced mean free path when applying magnetic fields. However, a quadratic dependence of resistance on magnetic field was predicted in this case and that clearly does not fit with the observed linear tendency of the MR [37]. Therefore, the magneto-transport in MoSe$_2$ layer is mainly attributed to the VRH process via localized states associated with the enhancement of the hopping rate, giving rise to a negative MR.

## III. CONCLUSION

In conclusion, we have investigated the fabrication of a single layer and a bilayer of MoSe$_2$ on sapphire using MBE. We found that the MoSe$_2$ layers are homogeneous over the substrate dimensions. By combining in-situ RHEED, in-plane XRD and STEM, we have shown that the layers are of polycrystalline nature and composed of nanometer-sized monodomains with free rotational angles. XPS and Raman spectroscopy confirm their stoichiometry and the homogeneous character over a large scale. The insulating characteristic of layered MoSe$_2$ was pointed out and interpreted in the theoretical framework of VRH. An unexpected negative magnetoresistance was evidenced at high temperatures and high magnetic fields. In the low charge carrier density system, this transport nature probably stems from the effect of magnetic field on the hopping characteristic through localized states induced by vacancy defects. This finding has revealed the intrinsic transport properties of layered TMDs which are valid for both natural and artificial sources at sub-micro scale.

**ACKNOWLEDGMENTS**

The authors thank Vincent Mareau, Laurent Gonon and João Paulo Cosas Fernandes for their assistance and discussions with Raman spectroscopy. The authors acknowledge Nicolas Mollard for preparation of TEM samples and Lucien Notin for technical assistance with MBE. This work is supported by the CEA-project 2D FACTORY and by the Agence Nationale de la Recherche within the ANR MoS2ValleyControl and 2D Transformers contracts. The LANEF framework (No. ANR-10-LABX-51-01) is also acknowledged for its support with mutualized infrastructure.

Phys. Rev. B **40**, 8342 (1989); R. Abdia, A. E. Kaaouachi, A. Nafidi, G. Biskupski, and J. Hemine, Solid State Electron **53**, 469 (2009).

[36] H. Zhang, X.-L. Fan, Y. Yang, and P. Xiao, J. Alloys Compd. **635**, 307 (2015).

[37] X. Zhang, Q. Xue, and D. Zhu, Phys. Lett. A **320**, 471 (2004).**SUPPLEMENTAL INFORMATION**

**1. MBE growth and substrate preparation**

All samples were grown in a MBE chamber with a base pressure of about $5 \times 10^{-10}$ mbar which increased up to about $2 \times 10^{-8}$ mbar during the co-deposition of Mo and Se. The Mo and Se are respectively evaporated by an e-beam gun and a home built evaporation cell. The direct Se flux measured at the sample position with a retractable Bayard-Alpert gauge is about $2 \times 10^{-6}$ mbar. The deposition rate of the Mo measured with a quartz micro-balance was 1.5 Å.min$^{-1}$. The post-annealing was performed at 720 °C during 15 min under Se flux. The as-received sapphire (0001) substrates were first cleaned in ultrasonic bath using acetone and iso-propanol, then annealed in a quartz tube oven in air at 300 °C for 30 minutes to desorb the organic species from the surface. The substrates were then heated up to 1000°C for 1h to smoothen the surface. They were finally loaded in the MBE system and outgassed in the UHV chamber at 700 °C for 15 minutes just before the deposition of the MoSe$_2$. After the preparation process, the sapphire surface is expected to be chemically inert with an Oxygen-Aluminium plane polarity.

**2. Structural analysis**

X-ray diffraction analysis was performed with a SmartLab Rigaku diffractometer equipped with a Copper Rotating anode beam tube ($K_\alpha$ = 1.54 Å) and operated at 45 kV and 200 mA. A parabolic mirror and a parallel in-plane collimator of 0.5° of resolution was used in the primary optics and a second parallel collimator was used in the secondary side. A $K_\beta$ filter was used for all the measurements. A Nanoscope Atomic force microscopy was used in tapping mode to image the samples. Scanning transmission electron microscopy was operated at an accelerated voltage of 80-200 kV using a probe aberration corrected microscope (Titan Themis FEI). The ball-and-stick model is built with VESTA software.

**3. Spectroscopic characterization**

Raman measurement were done with a Horiba Raman set-up using a laser excitation source f 632 nm with spot size of 0.5 µm. The signal was collected by choosing a 1800 grooves/mm grating. XPS measurements were carried out with a PHI 5000 VersaProbe II spectrometer delivering a micro-focused (100 m), monochromatized Al-K$_\alpha$ radiation (1486.6 eV). The overall energy resolution (X-ray bandwidth and electron analyser broadening) was 0.4 eV.

**4. Electrical measurements**

For electrical measurements, it is well known that metallic contacts deposited on exfoliated TMDs exhibit a large Schottky barrier due to the large bandgap Eg of the TMDs layer and its ultra-thin thickness. When TMDs layers are undoped, it is extremely difficult to inject charge carriers in the layers. To circumvent this, we have fabricated in-situ Mo(5nm)/Pd(10 nm) contacts (2 mm x 3 mm) that were deposited directly on the as-grown MoSe2 layer using a movable mechanical mask underneath the sample in the UHV chamber. The sample was then annealed at 200 °C for few minutes to obtain a good quality of contacts. The Mo/Pd pads were connected to a chip carrier with silver paste and Au wires. The temperature and magnetic field dependent transport measurements were performed with input currents of 10 nA and 50 nA respectively.

I-V curve is shown in Fig. S1 indicating an Ohmic nature of the contacts. Figure S2 shows MR at different temperatures with the linear fitting window of MR curves. The inset is the logarithmic dependence of fitted slopes-MR against temperature that allows us to infer the relationship slope-MR $\propto T^{-0.46}$

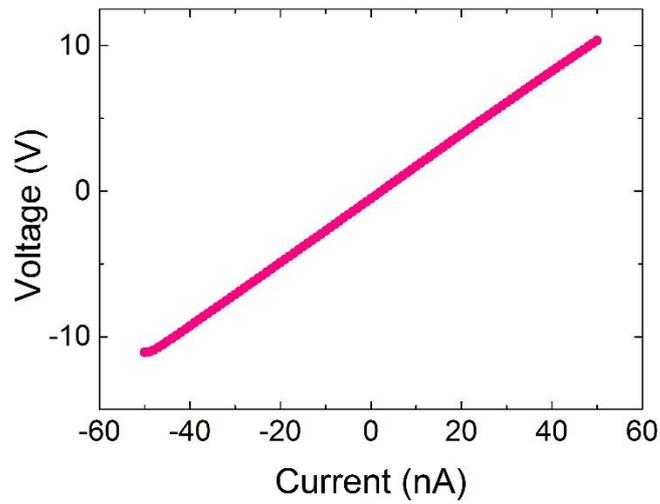

Fig. S1. I-V curve of the contact MoSe2/Mo/Pd measured at 250 K.

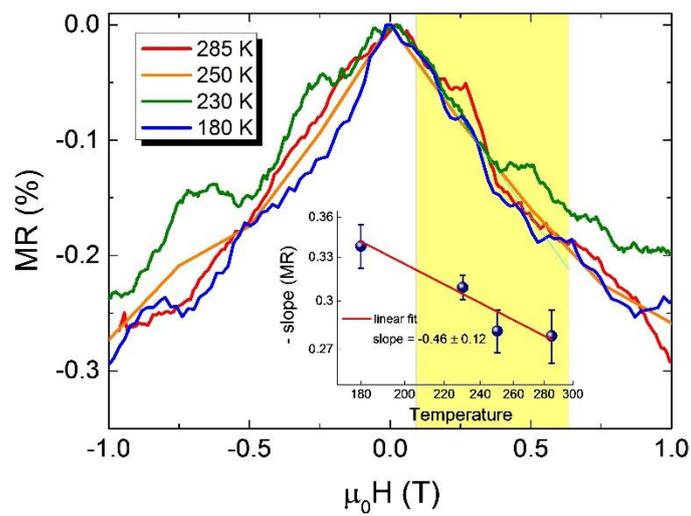

*Fig. S2.* Magnetoresistance of the bilayer at different temperatures. The yellow window represents the mediated-field region where the slope-MR was obtained. The inset is the plot of slope-MR versus temperature.